\documentclass[prb,twocolumn,showpacs,superscriptaddress,preprintnumbers,amssymb]{revtex4}
\usepackage{graphicx}
\usepackage{dcolumn}
\usepackage{bm}
\usepackage{wasysym}

\newcommand{\beq}{\begin{equation}}
\newcommand{\eeq}{\end{equation}}
\newcommand{\beqn}{\begin{eqnarray}}
\newcommand{\eeqn}{\end{eqnarray}}

\begin{document}
\title{Resonating plaquette phases in large spin cold atom systems}
\author{Cenke Xu}
\affiliation{Department of Physics, Harvard University, Cambridge,
MA 02138}
\author{Congjun Wu}
\affiliation{Department of Physics, University of California, San
Diego, CA 92093}
\date{\today}

\begin{abstract}
Large spin cold atom systems can exhibit novel magnetic properties
which do not appear in usual spin-1/2 systems. We investigate the
$SU(4)$ resonating plaquette state in the three dimensional cubic
optical lattice with spin-3/2 cold fermions. A novel gauge field
formalism is constructed to describe the Rokhsar-Kivelson type of
Hamiltonian and a duality transformation is used to study the
phase diagram. Due to the proliferation of topological defects,
the system is generally gapped for the whole phase diagram of the
quantum model, which agrees with the recent numerical studies. A
critical line is found for the classical plaquette system, which
also corresponds to a quantum many-body wavefunction in a
``plaquette liquid phase".
\end{abstract}
\maketitle

\section{Introduction}
Quantum fluctuations and non-Neel ordering magnetic states in low
dimensional spin-1/2 antiferromagnets are important topics in
strongly correlated physics. The quantum dimer model (QDM)
constructed by Rokhsar-Kivelson (RK) in which each dimer
represents an $SU(2)$ singlet provides a convenient way to
investigate novel quantum magnetic states such as the exotic
topological resonating valence bond (RVB) states
\cite{rokhsar1988}. The QDM in the 2d square lattice generally
exhibits crystalline ordered phase except at the RK point where
the ground state wavefunction is a superposition of all possible
dimer coverings \cite{fradkin1990}. In contrast, a spin liquid RVB
phase has been shown in the triangular lattice in a finite range
of interaction parameters by Moessner {\it et al}
\cite{moessner2001}. The 3d RVB type of spin liquid states have
also been studied by using the QDM \cite{huse2003, hermele2005}.

Recently, there is a considerable interest
on large spin magnetism with cold atoms in optical lattices
\cite{zhou2003,demler2002,wu2003,wu2006b,wu2005a,chen2005},
whose physics is fundamentally different from its counterpart in solid
state systems.
In solid state systems, the large spin on each site is formed by
electrons coupled by Hund's rule.
The corresponding magnetism is dominated by the exchange of a single
pair of spin-1/2 electrons, and thus quantum fluctuations are suppressed
by the large $S$ effect.
In contrast, it is a pair of large spin atoms that is exchanged in cold
atom systems, thus quantum fluctuations can even be stronger than those
in spin-1/2 systems.
In particular, a hidden and {\it generic} $Sp(4)$ symmetry has been
proved in spin-3/2 systems without of fine-tuning by Wu {\it et al.}
\cite{wu2003,wu2006b}.
This large symmetry enhances quantum fluctuations and
brings many novel magnetic physics \cite{wu2006b,wu2005,chen2005,
tu2006,tu2007}.

Below we will focus on a special case of spin-3/2 fermions at the
quarter-filling (one particle per site) in the 3D cubic lattice with
an $SU(4)$ symmetry which just means that all of the four spin
components are equivalent to each other.
The exchange model is the $SU(4)$ antiferromagnetic Heisenberg model with
each site in the fundamental representation.
Its key feature is that at least four-sites are required to form an
$SU(4)$ singlet two sites, i.e., two sites cannot form such a singlet.
This $SU(4)$ model was also constructed in spin-1/2 systems with orbital
degeneracy \cite{li1999,bossche2001}.
This model is different from the previous large-$N$ version of the $SU(N)$
Heisenberg model defined in the bipartite lattices where two neighboring
sites are with complex-conjugate representations and the $Sp(2N)$
Heisenberg model defined in non-bipartite lattice
\cite{arovas1988,sachdev1991}, both of which can have singlet dimers.
The natural counterpart of the dimer here is the $SU(4)$ singlet plaquette
state as $\frac{1}{4!} \epsilon_{\alpha\beta\gamma\delta}\psi^\dagger_\alpha(1)
\psi^\dagger_\beta(2) \psi^\dagger_\gamma(3) \psi^\dagger_\delta(4)$,
where $\alpha, \beta, \gamma,$ and $\delta$ take the value of $S_z$ as
$\pm\frac{3}{2}$ and $\pm\frac{1}{2}$.
Recently, the crystalline ordered $SU(4)$ plaquette state has been
investigated in quasi-1D ladder and 2D square lattice systems
\cite{bossche2000,bossche2001,chen2005}.
The resonating quantum plaquette model (QPM) in 3D has been constructed in
Ref. \cite{pankov2007} where quantum Monte-Carlo simulation shows that
the ground state is solid in the entire phase diagram.
The $SU(N)$ plaquette generalizations of the Affleck-Kennedy-Lieb-Tasaki
(AKLT) states \cite{affleck1987} have also been
given in Ref. \cite{arovas2007}.

In this article, we will formulate a novel gauge field
representation to the resonating plaquette model based on $SU(4)$
antiferromagnetic Heisenberg model in 3D cubic lattice. Unlike the
QDM in 3d cubic lattice, this QPM is generally gapped for the
whole phase diagram, due to the unavoidable proliferation of
topological defects. We study the novel gauge field in the dual
language, where a local description of topological defects is
possible. The classical ensemble of the plaquette system is also
discussed, and unlike its quantum version, our theory predicts the
classical ensemble can have an algebraic liquid phase by tuning
one parameter. Classification of topological sectors of the QPM is
also discussed.

\section{Quantum plaquette model}
\begin{figure}
\includegraphics[width=0.7\linewidth]{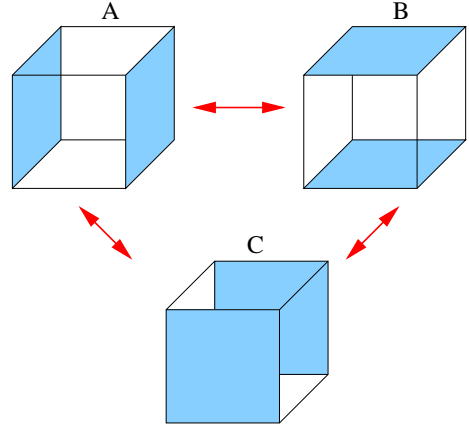}
\caption{Three flippable configurations in one cube.
The resonance is represented in the $t$-terim in Eq. \ref{RK}.
}
\label{fig:resonance}
\end{figure}

The QPM model in the 3D cubic lattice can be represented as
follows. The effective Hilbert space is constructed by all the
plaquette configurations allowed by the constraint: every site in
the cubic lattice is connected to one and only one plaquette.
Three flippable plaquette configurations exist in each unit cube
corresponding to the pairs of faces of left and right, top and
bottom, and front and back denoted as $A$, $B$ and $C$ in Fig.
\ref{fig:resonance}, respectively. The RK-type Hamiltonian
\cite{rokhsar1988} reads: \beqn
H&=&-t \sum_{\mbox{each cube}} \big\{ |A\rangle \langle B|+ |B\rangle \langle C|
+|B\rangle \langle C| +h.c.\big\} \nonumber\\
&+& V\sum_{\mbox{each cube}} \big\{ |A\rangle \langle A|  +
|B\rangle \langle B| + |C\rangle \langle C| \big\}, \label{RK}
\eeqn where $t$ has been shown to be positive in Ref.
\cite{pankov2007}, and we leave the value of $V/t$ arbitrary for
generality. Eq. \ref{RK} can be represented as \beqn H&=& t
\sum_{\mbox{each cube}} \big \{ |Q_1\rangle \langle Q_1|
+|Q_2\rangle \langle Q_2| \big \} + (V-2t) \nonumber \\ &\times&
\sum_{\mbox{each cube}} \big\{ |A\rangle \langle A|  + |B\rangle
\langle B + |C\rangle \langle C| \big\}, \eeqn where
$|Q_1\rangle=|A\rangle + \omega |B\rangle +\omega^2 |C\rangle$ ,
$|Q_2\rangle=|A\rangle + \omega^2 |B\rangle +\omega |C\rangle$,
and $\omega=e^{i\frac{2\pi}{3}}$. As a result, at $V = 2t$ (the RK
point), the ground state wavefunction should be annihilated by the
projectors $|Q_1\rangle\langle Q_1|$ and $|Q_2\rangle\langle
Q_2|$, i.e., the equal weight superposition between all the
plaquette configurations which can be connected to each other
through finite steps of local resonances, i.e., all the
configurations within one topological sector. At $V/t > 2$ all the
plaquette configurations without flippable cubes are eigenstates
of the Hamiltonian, one of which is the staggered plaquette state.
The phase diagram of this RK model has been studied numerically in
Ref. \cite{pankov2007}. In particular, the classical Monte Carlo
simulation performed shows that at this RK point a weak
crystalline order of resonating cubes is formed which forms a
cubic lattice with doubled lattice constant. At $V/t < 2$ the
system starts to favor flippable cubes. For instance, at $-V/t \gg
1$ the ground states are twelve fold degenerate with columnar
ordering. All the transitions between different phases are of the
first order.

The original RK Hamiltonian for quantum dimer model can be mapped
to the compact $U(1)$ gauge theory \cite{read1990,fradkin1990},
from which one can show that the 2+1 dimensional QDM is gapped
except for one special RK point, while 3+1 dimensional QDM has a
deconfined algebraic liquid phase \cite{sondhi2003}. By contrast,
the quantum plaquette model in the cubic lattice can be mapped
into a special type of lattice gauge field theory as follows. We
denote all  the square faces parallel to $XZ$ plane of the cubic
lattice by the sites to the left and bottom corner of the face: $i
+ \frac{1}{2}\hat{x} + \frac{1}{2} \hat{z}$, and denote faces
parallel to $XY$ and $YZ$ plane in a similar way. Then we define
the boson number $n$ with integer values on every face of the
cubic lattice. $n = 1$ corresponds to a face with plaquette, and
$n = 0$ otherwise. A strong local potential term $
U(n_{i+\frac{1}{2}\hat{\mu} + \frac{1}{2}\hat{\nu}} -
\frac{1}{2})^2$ is turned on at every face to guarantee the low
energy subspace of the boson Hilbert space is identical to the
Hilbert space with all the plaquette configurations. Since every
site is connected to one and only one plaquette, the summation of
$n$ over all twelve faces sharing one sites needs to be $1$. Next,
we define the rank-2 symmetric traceless tensor electric field on
the lattice as \beqn E_{i, \mu\nu} = \eta(i)
(n_{i+\frac{1}{2}\hat{\mu} + \frac{1}{2}\hat{\nu}} - \frac{1}{2}),
\ \ \ E_{\mu\nu} = E_{\nu\mu} (\mu \neq \nu), \label{definee}\eeqn
where $\eta(i) = (-1)^{i_x + i_y + i_z}$ equals $1$ when $i$
belongs to one of the two sublattices of the cubic lattice and
equals $-1$ otherwise. It is straightforward to check that the
one-site-one-plaquette local constraint on the Hilbert space can
be written compactly as \beqn \nabla_x \nabla_y E_{xy}+\nabla_y
\nabla_z E_{yz} + \nabla_z \nabla_x E_{zx}=5\eta(i),
\label{constraint}
\eeqn
where $\nabla$ is lattice derivative with
the usual definition $\nabla_\mu f = f(i+\hat{\mu}) - f(i)$.

The canonical conjugate variable of $E_{i,\mu\nu}$ is denoted as
the vector potential of $A_{i,\mu\nu}$, \beqn A_{i,\mu\nu} =
\eta(i) \ \theta_{i + \frac{1}{2}\hat{\mu} +
\frac{1}{2}\hat{\nu}}, \ A_{\mu\nu} = A_{\nu\mu}, \ (\mu \neq
\nu). \eeqn $\theta_{i + \frac{1}{2}\hat{\mu} +
\frac{1}{2}\hat{\nu}}$ is the canonical conjugate variable of
boson number $n_{i + \frac{1}{2}\hat{\mu} +
\frac{1}{2}\hat{\nu}}$, which is also the phase angle of boson
creation operator. $A_{\mu\nu}$ and $E_{\mu\nu}$ satisfy
\begin{eqnarray}
[E_{i,\mu\nu}, A_{j,\rho\sigma} ] &=& i\delta_{ij}
(\delta_{\mu\rho}\delta_{\nu\sigma} +
\delta_{\mu\sigma}\delta_{\nu\rho} ).
\label{eq:EA}
\end{eqnarray}
Because $E_{\mu\nu}$ only takes values with an integer step,
$A_{\mu\nu}$ is an compact field with period of $2\pi$. Due to the
commutator \beqn [E_{i,\mu\nu}, \exp (i A_{j,\nu\sigma}) ]=
(\delta_{\mu\rho}\delta_{\nu\sigma} +
\delta_{\mu\sigma}\delta_{\nu\rho} ) \exp( i A_{j,\nu\sigma}),
\eeqn operators $\exp(i A_{j,\nu\sigma})$ changes the eigenvalue
of $E_{i,\mu\nu}$ by 1. As a result, the plaquette flipping
process can be represented as \beqn H_t = &-& t\big\{\cos(\nabla_z
A_{xy} - \nabla_xA_{yz}) \\
&+& \cos(\nabla_x A_{yz} - \nabla_yA_{zx}) \nonumber \\
&+& \cos(\nabla_y A_{zx} - \nabla_zA_{xy}) \big\}, \label{low1}
\eeqn which is invariant under the gauge transformation of \beqn
A_{\mu\nu} \rightarrow A_{\mu\nu} + \nabla_\mu\nabla_\nu f,
\label{eq:gauge} \eeqn which is already implied by the local
constraint (\ref{constraint}). $f$ is an arbitrary scalar
function.
The low energy Hamiltonian of the system can be written as
\beqn
H &=& H_t + U\sum_{\mbox{each cube}} (E_{xy}^2+E_{yz}^2+E_{zx}^2)
\nonumber\\
&+&V \sum_{\mbox{each cube}} \Big\{ (\nabla_x E_{yz})^2 +
(\nabla_y E_{zx})^2+ (\nabla_z E_{zx})^2\Big\}, \nonumber \\
\label{low}
\eeqn
which is  subject to the constraint in Eq. \ref{constraint}.
Besides the gauge symmetry (\ref{eq:gauge}), Hamiltonian
(\ref{low}) together with constraint (\ref{constraint}) share
another symmetry as follows: \beqn \mu \rightarrow - \mu, \ \rho \rightarrow \rho, \ \sigma \rightarrow \sigma,  \\
E_{\mu\nu} \rightarrow - E_{\mu\nu}, \ E_{\sigma\mu} \rightarrow -
E_{\sigma\mu}, \ E_{\nu\sigma} \rightarrow  E_{\nu\sigma}, \\
A_{\mu\nu} \rightarrow - A_{\mu\nu}, \ A_{\sigma\mu} \rightarrow -
A_{\sigma\mu}, \ A_{\nu\sigma} \rightarrow  A_{\nu\sigma}.
\label{adsymmetry} \eeqn $\mu$, $\nu$ and $\sigma$ are three space
coordinates. This symmetry forbids terms like $E_{xy}E_{yz}$ to be
generated under RG flow at low energy.

\section{Duality transformation}

A major question in which we are interested is whether the
Hamiltonian Eq. \ref{RK} and Eq. \ref{low} have an intrinsic
liquid phase, just like the 3d QDM in the cubic lattice
\cite{sondhi2003}. A liquid state here corresponds to a gapless
Gaussian state in which we are allowed to expand the cosine
functions in Eq. \ref{low} at their minima, i.e., a ``spin wave"
treatment. However, the Gaussian phase could also be a superfluid
phase which breaks the conservation of boson numbers (or
effectively the plaquette numbers) with $\langle \exp(i\theta)
\rangle \neq 0$. In our current problem a superfluid phase is not
possible because $\langle \exp(i\theta) \rangle \neq 0$
necessarily breaks the local gauge symmetry (\ref{eq:gauge}) of
Hamiltonian (\ref{low}). In other words, a superfluid state is a
coherent state of boson phase $\theta$ implying a strong
fluctuation of boson numbers, which obviously violates the local
one-site-one-plaquette constraint.

In this type of lattice bosonic models, because bosonic phase
variable $A_{\mu\nu}$ is compact, the biggest obstacle of liquid
phase is the proliferation of topological defect, which tunnels
between two minima of the cosine function in Eq. \ref{low1}. Since
the topological defects are nonlocal, the best way to study them
is go to the dual picture, in which the topological defects become
local vertex operators of the dual height variables. similar
duality transformations have been used widely in studying all
types of bosonic rotor models, such as in proving the intrinsic
gap of 2D QDM \cite{fradkin1990,fradkin2004}, showing the
existence of ``bose metal phase" \cite{paramekanti} as well as the
deconfine phase of 3d QDM \cite{sondhi2003}, and very recently the
stable liquid phase of three dimensional ``graviton" model
\cite{xu2006c}.


\begin{figure}
\includegraphics[width=0.5\linewidth]{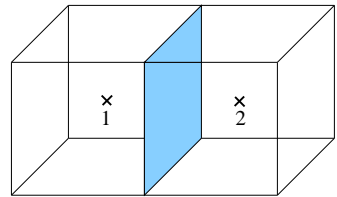}
\caption{The duality transformation defined in Eq. \ref{dual1}. On
dual site $1$ and $2$ there are three components of dual vector
height $h_\mu$, and the dual transformation for the shaded face is
$E_{yz} = (h_y - h_z)_2 - (h_y - h_z)_1 = \nabla_{x}(h_y - h_z)$.}
\label{dualfig}
\end{figure}

Besides the topological defects, another convenience one gains
from the dual formalism is the solution of the constraint, i.e.,
we are no longer dealing with a Hilbert space with a strict
one-site-one-plaquette constraint in Eq. \ref{constraint}. The
dual variables are defined on the dual lattice sites $\bar{i}$,
which are the centers of the unit cubes. In order to solve the
constraint
completely, one needs to introduce three components of the height
field $h_\mu$ ($\mu = 1$, 2, 3) on every dual site $\bar{i}$,
which is the center of a unit cubic of the original lattice: \beqn
E_{xy} &=& \nabla_{z}(h_{x} - h_{y})+E_{xy}^0, \nonumber \\
E_{yz} &=& \nabla_{x}(h_{y} - h_{z})+E_{yz}^0,  \nonumber \\
E_{zx} &=& \nabla_{y}(h_{z} - h_{x})+E_{zx}^0,
\label{dual1}
\eeqn
whose geometric illustration is shown in Fig. \ref{dualfig}.
$h_{x,y,z}$ are fields only take discrete integer vaules.
$E^0_{xy}$,$E^0_{yz}$ and $E^0_{zx}$ are background charges satisfying the
constraint Eq. \ref{constraint}.
We can just take the configuration of the columnar phase to define
the value of the background charges as
\beqn
E^0_{xy}(i,j,k)&=&  \left \{ \begin{array} {l} \frac{(-)^k}{2}
\ \ \ (\mbox{when both i and j are even})  \\
\frac{(-)^{i+j+k+1}}{2} \ \ \ (\mbox{otherwise})
                                \end{array} \right. \nonumber \\
E^0_{yz}(i,j,k)&=&E^0_{zx}(i,j,k)=\frac{(-)^{i+j+k+1}}{2}
\eeqn

The canonical momenta $\pi_{\mu}$ to the dual fields $h_{\mu}$
on each dual site are
\beqn \pi_x &=&  \nabla_y A_{zx} - \nabla_zA_{xy}, \\
\pi_y &=& \nabla_z A_{xy} - \nabla_xA_{yz},  \nonumber \\
\pi_z &=& \nabla_x A_{yz} - \nabla_yA_{zx}. \label{dual2} \eeqn
One can check the commutation relation and see that $\pi_\mu$ and
$h_\mu$ are a pair of conjugate variables. Then the dual
Hamiltonian of (\ref{low}) reads \beqn H& =& \sum_{\mu = x,y,z} -
t \cos \pi_\mu + U\sum_{\mu\nu\rho}
\zeta_{\mu\nu\rho} [\nabla_\mu(h_{\nu} - h_{\rho})-E^0_{\nu\rho}]^2\nonumber \\
&+&V \sum_{\mu\nu\rho} \zeta_{\mu\nu\rho} [\nabla_\mu (\nabla_\mu
(h_{\nu} - h_{\rho})-E^0_{\nu\rho})]^2 \label{dualhamil} \eeqn
$\zeta_{\mu\nu\rho}$ is a fully symmetric rank-3 tensor which
equals zero when any two of its three coordinates equal, and
equals one otherwise. On each dual lattice site $\bar{i}$, the
$\pi_\mu$ fields satisfy the relation
that $\sum_{\mu = x,y,z} \pi_{\mu, \bar{i}} = 0$.

The symmetry transformations of Hamiltonian Eq. \ref{dualhamil}
can be extracted from the duality transformation Eq. \ref{dual1}
and Eq. \ref{dual2}: \beqn h_x \rightarrow h_x + f(x,y,z) +
g_1(x), \cr h_y \rightarrow h_y + f(x,y,z) + g_2(y), \cr h_z
\rightarrow h_z + f(x,y,z) + g_3(z), \label{symmetry} \eeqn where
$f$ is a function of three spatial coordinates and $g_{1,2,3}$
only depends on one spatial coordinate. This type of symmetry is a
quasilocal symmetry, which also exists in the Bose metal states
\cite{paramekanti} and $p$-band cold atom systems \cite{xufisher}.



The main purpose of this paper is to study whether Hamiltonian
(\ref{dualhamil}) and (\ref{low}) have a liquid phase which
preserves all the lattice symmetries, just like the deconfined
algebraic liquid phase of 3D QDM. In this kind of algebraic liquid
phase, one can expand the cosine functions in equation (\ref{low})
and relax the discrete values of the $h_\mu$ fields, the long
distance physics can be described by a field theory which only
involves coarsed grained mode of $h_\mu$, let us denote the long
scale mode as $\tilde{h}_\mu$. In this Gaussian phase one can also
define continuous tensor electric field $\tilde{E}_{\mu\nu}$ as
the coarse grained mode of $E_{\mu\nu}$, the relation between
$\tilde{E}_{\mu\nu}$ and $\tilde{h}_\mu$ is $\tilde{E}_{\mu\nu} =
\zeta_{\mu\nu\rho}\partial_\rho (\tilde{h}_\mu - \tilde{h}_\nu) $.
A Gaussian field theory of $ \tilde{h}_\mu$ should satisfy the
continuous version of symmetries listed in Eq. \ref{symmetry} :
$\tilde{h}_\mu \rightarrow \tilde{h}_\mu + \tilde{f}(x,y,z) +
\tilde{g}_\mu(r_\mu)$, now $\tilde{h}_\mu$ as well as functions
$\tilde{f}$ and $\tilde{g}_\mu$ can all take continuous values. A
low energy field theory action is conjectured to be \beqn L =
\sum_{\mu} \frac{K}{2}(\partial_\tau \tilde{h}_\mu)^2 +
\frac{K}{2}\sum_{\mu\nu\rho} \zeta_{\mu\nu\rho}
(\nabla_\mu(\tilde{h}_{\nu} - \tilde{h}_{\rho}))^2 + \cdots,
\label{action} \cr \eeqn where the $\tilde{h}_{x,y,z}$ fields take
continuous real values. No other quadratic terms of
$\tilde{h}_\mu$ with second spatial derivative is allowed by
symmetry in this action.
Notice that in Eq. \ref{action} we have rescaled the
space time coordinates to make the coefficients of the first and
second term equal. The action (\ref{action}) describes a state
with enlarged conservation laws of $\pi_\mu$. If there is a state
described by the Gaussian action (\ref{action}), $\pi_x$, $\pi_y$
and $\pi_z$ are conserved within each $YZ$, $ZX$ and $XY$ plane
respectively. So any operator with nonzero expectation values at
this state has to satisfy the special 2d planar conservation law
of $\pi_\mu$.

The Gaussian part of action (\ref{action}) has one unphysical pure
gauge mode which corresponds to function $f$ in Eq.
\ref{symmetry}, and two gapless physical modes, with low energy
dispersion : \beqn \omega^2_{1} \sim k_x^2 + k_y^2 + k_z^2 +
\cr\cr \sqrt{k_x^4 + k_y^4 + k_z^4 - k_x^2k_y^2 - k_y^2k_z^2 -
k_x^2k_z^2}, \cr\cr \omega^2_{2} \sim k_x^2 + k_y^2 + k_z^2 -
\cr\cr \sqrt{k_x^4 + k_y^4 + k_z^4 - k_x^2k_y^2 - k_y^2k_z^2 -
k_x^2k_z^2}. \eeqn The second mode $\omega_2$ vanishes at every
coordinate axis of reciprocal space $(k_x, k_y, k_z)$. The strong
directional nature of $\omega_2$ roots directly in the quasilocal
gauge symmetries in Eq. \ref{symmetry}. The same modes can be
obtained from the continuum Gaussian limit action of Hamiltonian
(\ref{low}): \beqn L = \sum_{\mu \neq \nu}\frac{1}{2K}\{
(\partial_\tau \tilde{A}_{\mu\nu})^2 -
\zeta_{\mu\nu\rho}(\partial_\mu \tilde{A}_{\nu\rho} -
\partial_\nu \tilde{A}_{\rho\mu})^2 \}.  \eeqn In this action $\tilde{A}_{\mu\nu}$ is
the coarse grained mode of $A_{\mu\nu}$, and $\tilde{A}_{\mu\nu}$
is no longer a compactified quantity. The fact that $\omega_2$
vanishes at every coordinate axis plays very important role in our
following analysis, since it will create infrared divergence along
each axis in the momentum space, instead of only at the origin.
Similar directional modes are also found in other systems with
quasilocal symmetries \cite{paramekanti,xufisher}.

The ellipses in Eq. \ref{action} contain the non-Gaussian vertex
operators denoted as $L_v$, which manifests the discrete nature of
$h_\mu$. Since $h_\mu$ only takes integer values, a periodic
potential $\cos(2\pi h_\mu)$ can be turned on in the dual lattice
Hamiltonian (\ref{dualhamil}). At low energy the Non-Gaussian term
$L_v$ generated by $\cos(2\pi h_\mu)$ has to satisfy all the
symmetries in Eq. \ref{symmetry}, the simplest form it can take is
$\cos[2\pi \tilde{h}_\mu]$. However, this vertex operator only has
lattice scale correlation at the Gaussian fixed point, because it
violates the gauge symmetry of action (\ref{action}). Thus the
simplest vertex operator with possible long range correlation is
\beqn L_v = \sum_{\mu \neq \nu} - \alpha\cos[2\pi(\tilde{h}_\mu -
\tilde{h}_\nu)+\mathcal{B}_{\mu\nu}(\bar{i})]. \label{vertex}
\eeqn and $\mathcal{B}(\bar{i})$ is a function of dual sites,
which is interpreted as the Berry's phase. The specific form of
the Berry's phase of the vertex operators depends on the
background charge of the original gauge field formalism, which
determines the crystalline pattern of the gapped phase
\cite{fradkin2004}. However, since the liquid phase is a phase in
which the vertex operators are irrelevant, whether a liquid phase
exists or not does not depend on the Berry's phase, thus in the
current work we will not give a complete analysis of the Berry's
phase of our problem. In the continuum limit the most relevant
vertex operators are the ones with multi-defects processes without
Berry's phase and consistent with symmetries (\ref{symmetry}):
$\cos[2\pi N (\tilde{h}_\mu - \tilde{h}_\nu)]$, let us denote this
vertex operators as $ V_{N, \mu\nu}$, and integer $N$ can be
determined from detailed analysis of the Berry's phase. The
correlation function between two vertex operators with arbitrary
$N$ separated in space time is calculated as follows:
\begin{eqnarray}
\langle V_{N,\mu\nu}(0) V_{N,\mu\nu}(r) \rangle \cr\cr \sim \exp\{
- (2\pi)^2 N^2 \langle(\tilde{h}_\mu(0) - \tilde{h}_\nu(0)) \
(\tilde{h}_\mu(r) - \tilde{h}_\nu(r)) \rangle \} = \cr\cr
\delta_{r_\mu}\delta_{r_\nu}\exp\{ - \frac{2(2\pi)^2 N^2}{K}\int
\frac{d^4k}{(2\pi)^4} \frac{(2k_0^2 + 3k_\mu^2 +
3k_\nu^2)e^{i\vec{k}\cdot \vec{r}}}{(k^2_0 + \omega_1^2)(k^2_0 +
\omega_2^2)}
 \} \cr\cr\cr \rightarrow \delta_{r_\mu}\delta_{r_\nu} \mathrm{Const}
\ \  (r \rightarrow +\infty). \label{corre} \end{eqnarray} The
correlation function $\langle \tilde{h}(r)\
\tilde{h}(r^\prime)\rangle$ is evaluated at the Gaussian fixed
point described by the continuum limit action (\ref{action})
without $L_v$. The delta function $\delta_{r_\mu}\delta_{r_\nu}$
in Eq. \ref{corre} is due to the continuous quasilocal symmetry of
action (\ref{action}), or in other words the conservation of
$\pi_\mu$ within each planes. For instance, correlation function
$\langle e^{i 2\pi N \{\tilde{h}_x(0) - \tilde{h}_y(0)\} } e^{-i
2\pi N \{\tilde{h}_x (r)- \tilde{h}_y (r)\}} \rangle$ can only be
nonzero when $r_x = r_y = 0$, otherwise $\pi_x$ conservation
within every $YZ$ plane will be violated once $r_x \neq 0$.

Since the correlation function calculated in Eq. \ref{corre}
reaches a finite constant in the long distance limit, the vertex
operators are very relevant at the Gaussian fixed point described
by the action Eq. \ref{action}, and the system is generally gapped
with crystalline order in the whole phase diagram. Since this
result is applicable to any $N$ and independent of the Berry's
phase, the same conclusion is applicable to all the QPM with a
definite number of plaquette connected to each site. The specific
crystalline order can be determined from the detailed analysis of
the Berry's phase.

\section{Classical RK point}

At the RK point, the ground state wave-function is an equal weight
superposition of all the configurations allowed by constraint
(\ref{constraint}). All the static physics of this state is
mathematically equivalent to a classical ensemble, with partition
function defined as summation of all the plaquette configurations
with equal Boltzman weights. Since there is no energetic terms in
the partition function, all that rules is the entropy. If we
define the tensor electric field as Eq. \ref{definee}, the
classical ensemble can be written as \beqn Z = \sum_{E_{i,\mu\nu}}
\delta\{ \sum_{\mu \neq \nu} \ \nabla_\mu \nabla_\nu E_{\mu\nu} -
5\eta(i)\} \\ \times \exp\{ - U \sum_i \sum_{\mu \neq \nu}
(E_{i,\mu\nu})^2\}. \eeqn The delta function enforces the
constraint, and the term $- U \sum_{\mu \neq \nu}
(E_{i,\mu\nu})^2$ in the exponential makes sure all the low energy
$E_{\mu\nu}$ configurations are one-to-one mapping of the
plaquette configurations. Now solving the constraint by
introducing dual height field $h_{\mu}$, the classical partition
function can be rewritten as \beqn Z = \sum_{h_{\bar{i}, \mu}}
\exp \{ -
 U \sum_{\bar{i}} \sum_{\mu\nu\rho} \zeta_{\mu\nu\rho}
[\nabla_\mu(h_{\bar{i}, \nu} - h_{\bar{i},\rho})-E^0_{\nu\rho}]^2
\}. \eeqn

Again we are mainly interested in whether this classical ensemble
is an algebraic liquid state, or by tuning parameters one can
reach an algebraic liquid phase. We can conjecture a low energy
classical field theory generated by entropy, allowed by symmetry
(\ref{symmetry}). The same strategy has been used to study
classical six-vertex model, classical three-color model and four
color model \cite{kondev1996}. Here the simplest low energy
effective classical field theory reads \beqn F = \sum_{\mu\nu\rho}
\frac{\tilde{K}}{2} \zeta_{\mu\nu\rho} (\nabla_\mu(\tilde{h}_{\nu}
- \tilde{h}_{\rho}))^2 + \cdots. \label{classical}\eeqn The number
$\tilde{K}$ cannot be determined from our field theory. This is
the simplest free energy allowed by symmetry. The physical meaning
of this free energy is that, the total number of plaquette
configurations (entropy) in a three dimension volume is larger if
the average tensor electric field $E_{\mu\nu}$ is small, i.e. the
entropy favors zero average tensor electric field.

The ellipses in equation (\ref{classical}) includes the vertex
operators in equation (\ref{vertex}). The relevance of the vertex
operators can be checked by calculating the scaling dimensions of
the vertex operators at the Gaussian fixed point action
(\ref{classical}). Let us denote vertex operator $\cos[2\pi
N(\tilde{h}_\mu - \tilde{h}_\nu)]$ as $V_{N, \mu\nu}$. Due to the
symmetry (\ref{symmetry}), $V_{N, xy}$ can only correlates with
itself along the same $\hat{z}$ axis, and $V_{N, zx}$ and $V_{N,
yz}$ can never have nonzero correlation between each other when
they are separated spatially along $\hat{z}$ axis.

The leading order correlation functions are \beqn \langle V_{N,
xy, (0,0,0)}V_{N,xy,(0,0,z)}\rangle \cr\cr \sim \exp\{ -
\frac{(2\pi)^2 N^2 2}{\tilde{K}}\int
\frac{d^3k}{(2\pi)^3}\frac{(k_x^2 + k_y^2) e^{ik_z z}}{k_x^2k_y^2
+ k_y^2k_z^2 + k_z^2k_x^2} \} \cr\cr = \frac{1}{z^{8\pi
N^2/\tilde{K}}}. \eeqn In the above calculations we have chosen
the simplest regularization: replacing spatial derivative on the
lattice by momentum $ik_x$. It has been shown that the scaling
dimensions of operators in these type of models with extreme
anisotropy can depend on the regularization on the lattice
\cite{paramekanti}. Here the scaling dimension of operator
$V_{N,\mu\nu}$ is regularization independent. These vertex
operators are irrelevant if $\tilde{K} < \tilde{K}_c = 4\pi N^2$,
in this parameter regime the contribution of $V_{N, xy}$ to
various correlation functions can be calculated perturbatively.

Some other vertex operators can be generated under renormalization
group flow, but these vertex operators all have algebraic
correlations, with a regularization dependent scaling dimension
proportional to $1/\tilde{K}$. For instance, vertex operator
$\cos[2\pi N\nabla^n_{mx}(\tilde{h}_x - \tilde{h}_y)]$ has nonzero
algebraic correlation function in the $YZ$ plane at long distance.
Here lattice derivative $\nabla_{mx}$ is defined as $\nabla_{mx}
f(\vec{r}) = f(\vec{r} + m\hat{x}) - f(\vec{r})$. If we regularize
the theory by replacing lattice derivative $\nabla_{mx}$ with $i
2\sin(mk_x/2)$ in the momentum space, the scaling dimension of
$V_{N,mx,xy} = \cos[2\pi N\nabla^n_{mx}(\tilde{h}_x -
\tilde{h}_y)]$ with $ n = 1$ and arbitrary integer $m$ is $ 4\pi
N^2/\tilde{K}$, and the scaling dimension is isotropic in the
whole $YZ$ plane: \beqn \langle
V_{N,mx,xy,(0,0,0)}V_{N,mx,xy,(0,y,z)} \rangle \sim
\frac{c(\theta)}{(y^2+z^2)^{4\pi N^2/\tilde{K}}}. \eeqn Here
$c(\theta)$ is a positive function of $\theta = \arctan(z/y)$.
Notice that the rotation symmetry in the $YZ$ plane is not
restored even at long length scale. The scaling dimension of
$\cos[2\pi N\nabla^n_{mx}(\tilde{h}_x - \tilde{h}_y)]$ increases
rapidly with number $n$. Thus all the vertex operators are
irrelevant when $\tilde{K}$ is small enough, and there is a
critical $\tilde{K}_c$ separating a crystalline order and the
algebraic liquid phase. At the liquid line the crystalline order
parameter should have algebraic correlation functions. Coefficient
$\tilde{K}$ can be tuned from adding energetic terms in the
system. Recall that now the configurations with zero average
$E_{\mu\nu}$ are favored by entropy, if we want to reduce
$\tilde{K}$, we can add energetic terms which disfavor zero
average $E_{\mu\nu}$. For instance, if we give the flippable
cubics a smaller weight than the unflippable cubics, coefficient
$\tilde{K}$ should be reduced.

The above results can be roughly understood from simple physical
argument. Notice that all the flippable cubes have zero average
electric field, so the entropy effectively favors flippable cubes.
If $\tilde{K} > \tilde{K}_c$, the entropy strongly favors
flippable cubes, the system will develop crystalline order which
maximize the number of flippable cubes. This kind of effect is
usually called ``order by disorder". It is also natural that the
crystalline order tends to be weakened or even melt if we reduce
$K$. Since the melting transition of the crystalline order is
driven by the proliferation of defect operators, the universality
class of this transition is very similar to the
Kosterlitz-Thouless transition of the 2D XY model, the correlation
length between operators $V_{N,\mu\nu}$ diverges as $ \xi \sim
\exp[\sqrt{b/(\tilde{K} - \tilde{K}_c)}]$. Unusual KT like
transition in 3d or higher dimensions have also been discussed in
other systems with similar quasilocal symmetries
\cite{paramekanti}, where the dimensionality of the system is
effectively reduced to 2d.

Recent Monte Carlo simulation \cite{pankov2007} shows that the
whole phase diagram of RK Hamiltonian (\ref{RK}) is gapped with
crystalline order, including the RK point. Our results based on
duality is consistent with this numerical results, and the equal
weight classical partition function should have $\tilde{K} >
\tilde{K}_c$. Our theory also predicts that if we turn on
energetic terms which favors unflippable cubes, there is a
critical line described by Gaussian field theory
(\ref{classical}). This prediction can be checked by classical
Monte Carlo simulations. Another prediction which in principle can
be made in our formalism is the most favored crystalline order
when $\tilde{K}$ is slightly larger than $\tilde{K}_c$. This
requires a detailed analysis of the Berry's phase of the vertex
operators in the dual theory, which we leave to future studies.

\section{topological sector}

\begin{figure}
\includegraphics[width=0.9\linewidth]{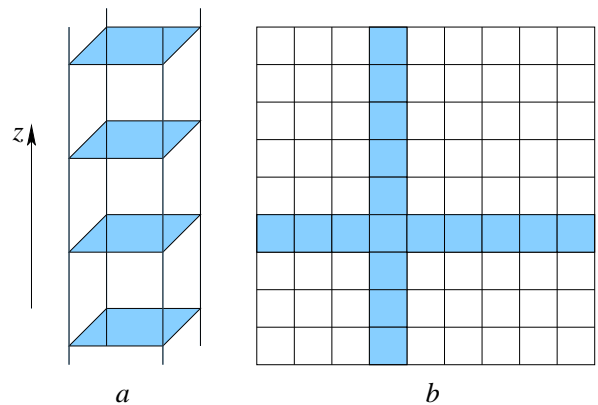}
\caption{($a$), the conserved quantity $m_{xy}$ is defined as the
summation of all $E_{xy}$ on all the shaded squares along one $z$
axis. ($b$), the view of the 3d lattice from the top. If the
quantity $m_{xy}$ is fixed on all the shaded squares shown in this
figure, $m_{xy}$ is determined on the whole lattice. }
\label{topo2}
\end{figure}

Now let us discuss the topological sector, within which every
configuration can be connected to each other through finite local
movings depicted in Fig. \ref{fig:resonance}. Topological sectors
are especially useful when one is dealing with a quantum liquid
state, where Landau's classification of phases are no longer
applicable. In the original quantum dimer model on square lattice,
the topological sector on a torus is specified by two integers
\cite{rokhsar1988}, which can be interpreted as winding numbers of
electric fields. Here we choose a lattice with even number of
sites in each axis and impose the periodic boundary condition. To
specify a topological sector one needs to know the conserved
quantities under local movings. It is straightforward to check
that quantity $m_{i_x,i_y, xy} = \sum_{i_z} E_{i, xy}$ for any 2d
coordinate $(i_x, i_y)$ is a conserved quantity. Notation
$\sum_{i_z}$ means summation over all the sites with the same $x$
and $y$ coordinates (Fig. \ref{topo2}). However, these quantities
are not independent. For instance, using constraint
(\ref{constraint}) we have the following identity: \beqn
m_{0,0,xy} &-& m_{1,0,xy} + m_{1,1,xy} - m_{0,1,xy}  = \sum_{i_z}
\nabla_x \nabla_y E_{xy} \cr\cr &=& \sum_{i_z} 5 \eta(i) - (
\nabla_y \nabla_z E_{yz} + \nabla_z \nabla_x E_{zx}) = 0. \eeqn
Thus as long as one fix the quantity $m_{xy}$ for one column and
one row in the $XY$ plane, their values for the whole lattice are
determined. Conserved quantities associated with $E_{zx}$ and
$E_{yz}$ can be treated in the same way. Thus we conclude that one
needs infinite number of integers to specify a topological sector
on a three dimension torus, and the number scales with the linear
size of the lattice.

\section{Summary and comparison with other models}

This work studies a three dimensional quantum resonating plaquette
model, motivated from a special SU(4) invariant point in spin-3/2
cold atom system. The effective low energy physics of the problem
can be mapped to a special type of lattice gauge field. Our
current QPM together with previously studied 3d QDM
\cite{sondhi2003} and soft-graviton model \cite{xu2006c} all have
local constraint and low energy gauge field description without
gapless matter fields. Unlike the QDM and the soft-graviton model,
the QPM almost always suffers from the proliferation of
topological defects, and a generic stable algebraic liquid state
as an analogue of the photon phase of 3d QDM does not exist.

The reason of the existence of a stable liquid phase of 3d QDM as
well as the 3d soft-graviton model have been discussed in
reference \cite{xu2006c}. Both models with stable liquid phases
are self-dual gauge theories, with strong enough gauge symmetries
in both the original description of the problem or the dual
theories, i.e. one cannot write down a gauge invariant vertex
operator which gaps out the liquid phase. In our current QPM, the
symmetry of the dual theory does not rule out all the vertex
operators, and gauge invariant vertex operators are very relevant.
Thus in this type of bosonic quantum rotor models, large enough
gauge symmetries are necessary for both sides of the duality to
guarantee the existence of a stable liquid phase if gapless matter
field is absent.

\begin{acknowledgments}
The authors thank D. Arovas, L. Balents, S. Kivelson and S. Sondhi
for helpful discussions. C. W. is supported by the start up
funding at University of California, San Diego; Cenke Xu is
supported by the Milton Funds of Harvard University.

\end{acknowledgments}

\bibliography{spin33,plaquette}
\end{document}